\documentclass[conference]{IEEEtran}
\IEEEoverridecommandlockouts
\usepackage{cite}
\usepackage{amsmath,amssymb,amsfonts}
\usepackage{algorithmic}
\usepackage{graphicx}
\usepackage{textcomp}
\usepackage{xcolor}
\usepackage[hyphens]{url}
\usepackage{comment}
\usepackage[breaklinks=true]{hyperref}

\def\BibTeX{{\rm B\kern-.05em{\sc i\kern-.025em b}\kern-.08em
    T\kern-.1667em\lower.7ex\hbox{E}\kern-.125emX}}
\begin{document}

\title{Unikernels vs. Containers: A Runtime-Level Performance Comparison for Resource-Constrained Edge Workloads}

\author{\IEEEauthorblockN{Hai Dinh-Tuan}
\IEEEauthorblockA{\textit{Service-Centric Networking} \\
\textit{Technische Universit\"{a}t Berlin}\\
Berlin, Germany \\
hai.dinhtuan@tu-berlin.de}
}
\maketitle

\begin{abstract}
The choice between containers and unikernels is a critical trade-off for edge applications, balancing the container's ecosystem maturity against unikernel's specialized efficiency. However, until now, how this trade-off behaves under the severe memory constraints of industrial edge environments remains insufficiently investigated, especially across different execution models. This work presents an empirical comparison using Go and Node.js applications, representing ahead-of-time (AOT) and just-in-time (JIT) compilation, respectively. While unikernels consistently deliver faster startup times and outperform containers for Go-based workloads in resource-constrained environments, the evaluation results identify a critical performance crossover for Node.js. Below a certain memory threshold, Docker containers maintain stable performance for both I/O-bound and CPU-bound applications, while the Nanos unikernel's performance degrades sharply. This reveals that Linux’s memory management capabilities can outweigh the minimalist efficiency of unikernels under resource scarcity, a critical trade-off that, until now, has not been adequately quantified for JIT runtimes in this context. These findings demonstrate that the optimal deployment paradigm depends on both runtime behavior and available system resources, underscoring the need for workload-aware deployment strategies in edge computing.

\end{abstract}


\section{Introduction}

With the growing adoption of cloud-native technologies across industries \cite{8780345}, new approaches are required for both \textit{service development} and \textit{service deployment}. While the development aspect has received considerable research attention \cite{10.1145/3393822.3432339}, deployment practices are still largely shaped by the dominance of container technologies. Yet, the choice of how services are packaged and deployed represents a critical architectural decision with direct implications for performance, security, and resource utilization. Therefore, this work aims to compare \textit{containers} and \textit{unikernels}, particularly in the context of edge workloads operating under resource constraints and across different programming language execution models.

\begin{figure}[htbp]
\centering
\includegraphics[width=\linewidth]{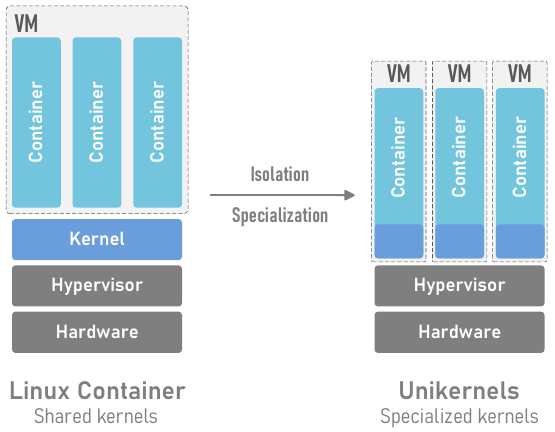}
\caption{Architectural comparison showing a Linux container's reliance on a shared host kernel versus a unikernel's specialized, self-contained design running on a hypervisor (recreated from \cite{arado2021unikernels}).}
\label{fig:container_unikernel}
\end{figure}

Containerization (OCI/Docker) is the de facto way to package and run applications today, offering a mature, robust, and developer-friendly ecosystem. By using Operating System (OS)-level features like Linux \textit{namespaces} and \textit{cgroups}, containers provide lightweight application isolation, simplifying deployment and aligning with established cloud-native practices. In contrast, unikernels represent a highly specialized approach by compiling an application with only the necessary OS libraries into a minimal, single-address-space machine image that runs directly on a hypervisor, as shown in Fig.~\ref{fig:container_unikernel}.

While the trade-off between containers and unikernels is extensively investigated in theory, its practical implications under the constrained and dynamic memory conditions typical of industrial edge environments remain underexplored, particularly for modern execution models. In particular, it is unclear how just-in-time (JIT) compiled runtimes perform in comparison to ahead-of-time (AOT) compiled languages. This work addresses this gap through a comparative benchmark of Docker containers and the Nanos unikernel platform, using both AOT (Go) and JIT (Node.js) applications under varying memory constraints. This work's contributions are therefore threefold:

\begin{itemize}
    \item \textit{Identification of a performance crossover point for JIT:} We show that unikernels perform better with sufficient resources, but below a certain threshold, Docker's mature Linux environment maintains stable performance, while unikernel latency and throughput degrade significantly.

    \item \textit{Linking execution model to deployment paradigm:} For AOT workloads, unikernels consistently outperform containers under resource-limited conditions. In contrast, for JIT workloads, Docker offers greater stability, especially under resource constraints.

    \item \textit{Decision framework for deployment choice:} We present practical guidance for selecting the appropriate deployment model based on workload type, execution model, and system constraints.
\end{itemize}

The remainder of this work is structured as follows: Section II discusses related work, Section III details our experimental methodology. Section IV presents our empirical results, and Section V discusses their implications. Finally, Section VI concludes the paper.

\section{Related work}

Docker containers have emerged as the \textit{de facto} industry standard, primarily due to their vast community and industry-wide adoption. The availability of robust orchestration tools like Kubernetes\footnote{\url{https://kubernetes.io/}} and Docker Swarm\footnote{\url{https://docs.docker.com/engine/swarm/}} provides robust frameworks for scaling, service discovery, and lifecycle management \cite{10942720}, which significantly simplifies the deployment of complex applications \cite{cito2017empirical, siddiqui2019comprehensive, 9766576, queiroz2023container}. However, this approach also comes with inherent overhead and a broader attack surface, as all containers on a host share the same underlying OS kernel.

Unikernels, by design, offer advantages in both security and efficiency. By compiling an application with only the necessary operating system libraries into a minimal, single-address-space machine image, unikernels offer a significantly reduced attack surface and hypervisor-level isolation \cite{chen2021evolving, combe2016docker, talbot2020security}. Their minimalist architecture also leads to minimal image sizes and boot times in the millisecond range, making them well-suited for resource-constrained and latency-sensitive scenarios such as the network edge \cite{goethals2022functional, chen2022unikernel}.

Several empirical studies have been proposed to quantify these differences, revealing that the performance outcomes are highly dependent on the specific workload, hardware architecture, and implementation context. For instance, Mavridis et al. \cite{mavridis2019lightweight} found that while Docker containers generally achieved a higher request rate, MirageOS unikernels offered superior performance with lower latency and minimal memory usage, albeit with the limitation of being language-specific (OCaml). 

Several studies have also compared containers and unikernels in specific use cases. In the context of 5G Virtual Network Functions (VNFs), Filipe et al. \cite{filipe2019performance} found that Docker containers had faster instantiation times and lower end-to-end latency, but the IncludeOS unikernel showed an 11\% higher throughput for high-bandwidth UDP traffic, highlighting a trade-off between control plane agility and data plane performance. For serverless platforms at the edge, Mistry et al. \cite{mistry2020demonstrating} demonstrated that a MirageOS-based serverless platform on a resource-constrained Raspberry Pi significantly outperformed a container-based OpenWhisk solution in processing speed, memory consumption, and boot latency, with startup times an order of magnitude faster.

However, the maturity of the unikernel ecosystem and its performance on different hardware architectures remain critical considerations. Acharya et al. \cite{acharya2018performance} highlighted the immaturity of unikernels on the ARMv8 architecture, where KVM was found to be the fastest for network I/O, in stark contrast to the x86 platform where containers and unikernels held an advantage. This underscores that performance is not guaranteed as a whole when using unikernels; rather, it depends strongly on the underlying hardware architecture and the maturity of the specific framework.

In addition, the majority of existing comparisons use either system-level benchmarks (e.g., Iperf, SysBench) \cite{acharya2018performance} or focus on applications written in AOT-compiled languages like C++ or OCaml \cite{filipe2019performance, mistry2020demonstrating}. A significant gap remains regarding systematic, controlled evaluations of how JIT-compiled runtimes like Node.js behave in a unikernel versus a container. Some studies do use resource-constrained hardware, such as Raspberry Pi \cite{mistry2020demonstrating}, but their conclusions are often oversimplified, typically claiming that one deployment model outperforms the other in that environment, without systematically varying constraints (particularly memory) to reveal why, or under what conditions, such advantages hold.

This research addresses that gap by presenting a comprehensive empirical evaluation that isolates the impact of deployment models (Docker vs. Nanos) across both AOT and JIT runtimes under controlled memory and CPU constraints. Unlike prior work, it quantifies how performance characteristics change across execution models and identifies the threshold at which one model outperforms the other. Furthermore, it considers a full spectrum of metrics, including latency, jitter, cold-start time, and memory footprint. By proposing a systematic analysis and translating results into a deployment decision framework, this study not only complements previous research but also provides a deeper and more comprehensive understanding of the trade-offs between unikernels and containers. For the study, we firstly hypothesize that:

\begin{itemize}
    \item \textbf{H1 (Static Properties):} Unikernels will demonstrate an order-of-magnitude advantage in startup time and a significantly smaller image footprint compared to containers, confirming the benefits of their minimalist architecture.
    \item \textbf{H2 (Dynamic Performance):} The performance trade-off will be workload-dependent. Unikernels will have an advantage in raw, single-core CPU-bound tasks due to minimal overhead, whereas Docker containers will outperform in I/O-bound, multi-core workloads by utilizing the host’s optimized kernel. For JIT runtimes under memory constraints, we further expect a performance crossover, where the stability of the mature Linux environment outweighs the unikernel’s efficiency.
\end{itemize}

\section{Experimental Design}

In this study, we designed a series of synthetic benchmarks to quantify the performance and resource trade-offs between Docker containers\footnote{\url{https://www.docker.com/}}  and Nanos unikernels\footnote{\url{https://nanos.org/}}. 

Two functionally identical applications were developed in Go (v1.22.2) and Node.js (v24.3.0) to compare an AOT-compiled language with a JIT-compiled language, reflecting differences in execution models under each deployment paradigm\footnote{The source code is available at: \url{https://github.com/haidinhtuan/Unikernel-vs-Container}}. Each application exposes two REST API endpoints:

\begin{itemize}
    \item \textit{I/O-Bound Workload (/io):} Returns a simple response, intended to evaluate networking and scheduling performance with minimal application overhead.
    \item \textit{CPU-Bound Workload (/compute):} Executes a 100-iteration SHA-256 hashing loop prior to generating a response, designed to stress computational performance.
\end{itemize}

Experiments were conducted on two dedicated servers located in the same data center. One hosted the application under evaluation, while the other generated test requests. This setup was designed to minimize cross-interference between the request generation and application processing. The configuration of the test server is as follows:

\begin{itemize}
    \item CPU: Intel Xeon E3-1275 v6 (4 cores, 8 threads, 3.80~GHz) with hardware-assisted virtualization enabled to eliminate the overhead of software-based emulation.
    \item Memory: 64~GB DDR4 ECC RAM.
    \item Network: 1~Gbps full-duplex.
    \item OS: Ubuntu 24.04.2 LTS, Linux kernel 6.8.0-57-generic.
    \item Container Runtime: Docker v28.3.1.
    \item Unikernel Runtime: Nanos v0.1.54 with ops v0.1.43 via QEMU v8.2.2.
\end{itemize}

The evaluation focuses on two main aspects:

\begin{itemize}
    \item \textit{Static performance:} Image footprint (MB) and time to ready (cold-start latency).
    \item \textit{Dynamic performance:} Throughput (req/s), latency (ms), and jitter (ms).
\end{itemize}

All of the test cases were conducted under two scenarios:
\begin{itemize}
    \item \textit{Resource-Rich Scenario:} 8 CPU cores, 2048~MB RAM, simulating a well-provisioned edge server.
    \item \textit{Resource-Constrained Scenario:} Single CPU core, memory varied across 512~MB, 256~MB, 128~MB, and 64~MB, simulating industrial edge devices.
\end{itemize}

\section{Results and analysis}

\subsection{Static Properties: Footprint and Startup Time}

\begin{figure}[htbp]
\centering
\includegraphics[width=\linewidth]{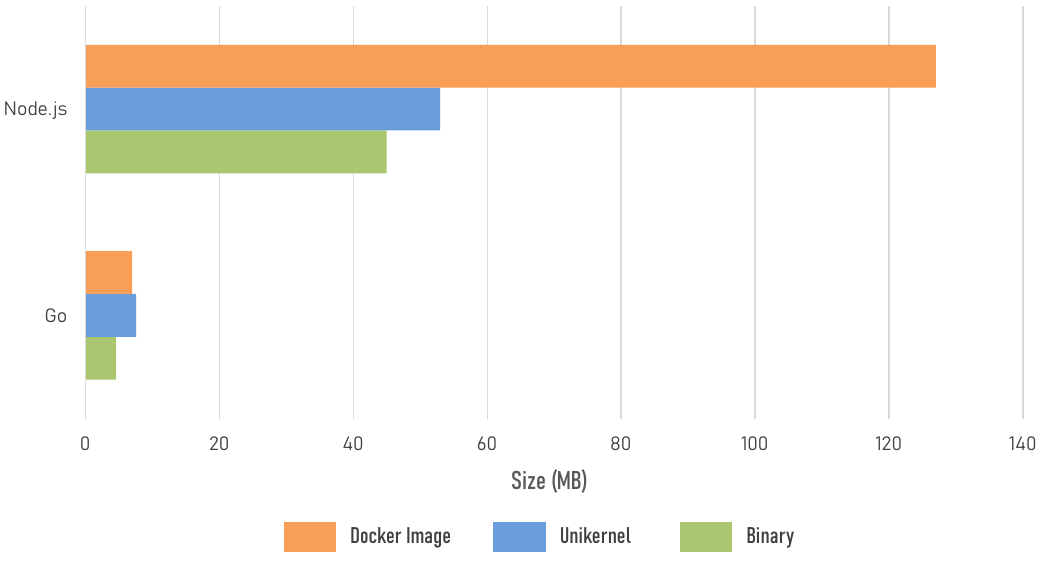}
\caption{Comparison of image size between container and unikernel.}
\label{fig:image_size}
\end{figure}

For the Go application, we compare the raw binary (4.6 MB), the Nanos image (7.6 MB), and a Docker image built using \texttt{golang:1.22-alpine} as the \textit{builder} and \texttt{scratch} as the \textit{base} (7.0 MB). The Docker image is slightly larger than the raw binary due to image metadata and layering, while the Nanos image is slightly larger than the Docker image because it includes the Nanos kernel (around 3 MB), as unikernels need to package their own kernel while containers share the host’s kernel. 

For the Node.js application, the raw binary created with \textit{pkg} is 45.0 MB. The Docker image (\texttt{FROM node:alpine}) expands to 127.0 MB due to the Alpine OS, the full Node.js runtime and V8 engine, and the application payload. In contrast, the Nanos image is only 53.0 MB, containing the application, the Nanos kernel, and minimal Node.js libraries required to run the app. Thus, for interpreted/JIT-compiled languages like Node.js, Nanos achieves a 2.4x size reduction over Docker by removing the guest OS layer as shown in Fig.~\ref{fig:image_size}.

\begin{figure}[htbp]
\centering
\includegraphics[width=0.8\columnwidth]{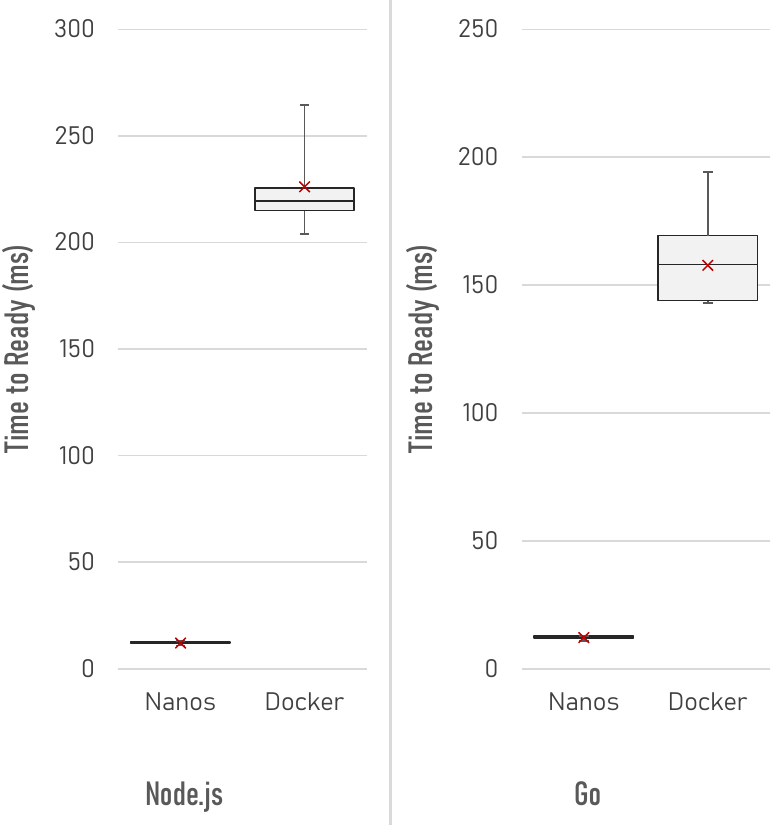}
\caption{Comparison of time-to-ready between container and unikernel.}
\label{fig:ttr}
\end{figure}

To quantify boot time, we use \textit{Time to Ready (TTR)} as the duration from issuing the launch command to the application serving its first network request. Across Go and Node.js, the unikernel consistently achieves a median TTR of 12 ms, while Docker containers require 158.00 ms (Go) and 219.50 ms (Node.js). The results are illustrated in Fig.~\ref{fig:ttr}, showing that the unikernel is 13x–18x faster to become operationally ready and demonstrates its highly deterministic and predictable boot process.

\subsection{Performance Evaluation: I/O-Bound Workload}

\subsubsection{Resource-Rich Scenario}

\begin{figure}[htbp]
\centering
\includegraphics[width=\linewidth]{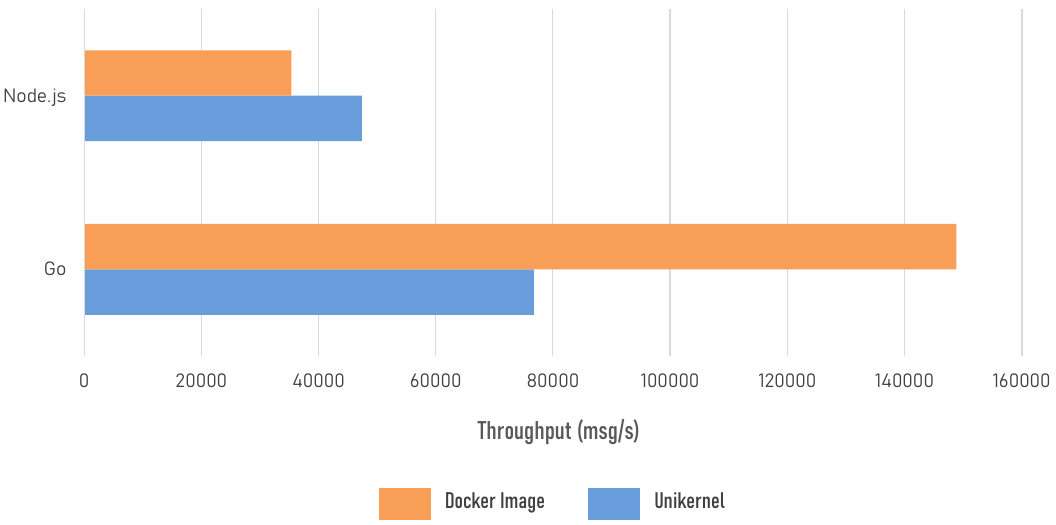}
\caption{Comparison of Throughput between container and unikernel with high resource allocation.}
\label{fig:io_throughput_unconstrained}
\end{figure}

Both Docker and Nanos were given 8 CPU cores and 2048 MB of memory to eliminate unrelated hardware constraints. For Go, Docker achieved 148,822.64 req/s, double that of Nanos (76,741.51 req/s) as shown in Fig.~\ref{fig:io_throughput_unconstrained}. This gap is due to Docker using the host’s optimized kernel-level networking, efficiently handling high parallel I/O, while Nanos, using user-mode networking by default via \textit{ops}, incurs higher overhead, limiting throughput.\footnote{\url{https://docs.ops.city/ops/networking} (accessed on Dec 6, 2024).} In contrast, for Node.js, Nanos outperformed Docker (47,368.19 req/s vs. 35,331.71 req/s). As Node.js is single-threaded by default, it cannot fully utilize all CPU cores, shifting the bottleneck from network capacity to runtime efficiency. Here, Nanos’s minimal, OS-free environment offers lower overhead, benefiting the busy \textit{event loop} and improving performance.

The Go application on Docker exhibited an average latency of 2.67ms with an exceptionally low standard deviation of 499.35µs, indicating very little jitter between requests. In contrast, the Go application on the Nanos unikernel, while still fast with an average latency of 4.12ms, showed a standard deviation of 47.58 ms, a value nearly two orders of magnitude higher than Docker's. This high level of jitter indicates a less predictable response time. Furthermore, the system dropped 37 requests due to timeouts, suggesting that while its average performance was good, it experienced moments of saturation or scheduling contention that led to failed requests. 

\subsubsection{Resource-Constrained Scenario}

\begin{figure}[htbp]
\centering
\includegraphics[width=\linewidth]{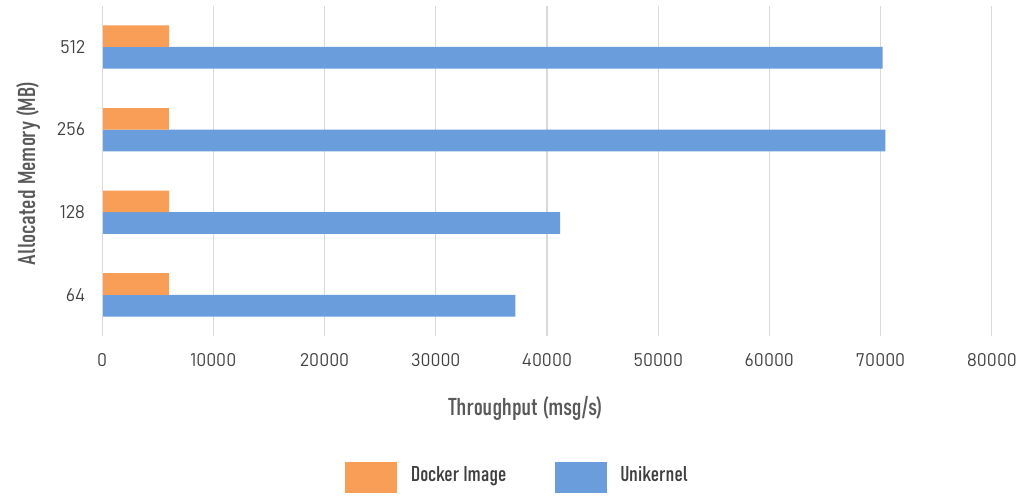}
\caption{I/O-Bound Throughput for the Go Application under Varying Memory Constraints (1 CPU). The Nanos unikernel consistently delivers an order-of-magnitude higher throughput.}
\label{fig:io_constrained_go_throughput}
\end{figure}

To simulate a typical resource-constrained industrial edge device, a second set of benchmarks was conducted in which both Docker and Nanos were restricted to a single CPU core. The available memory was then varied across 512 MB, 256 MB, 128 MB, and 64 MB to analyze each platform’s behavior under increasing memory constraints.

\begin{figure}[htbp]
\centering
\includegraphics[width=\linewidth]{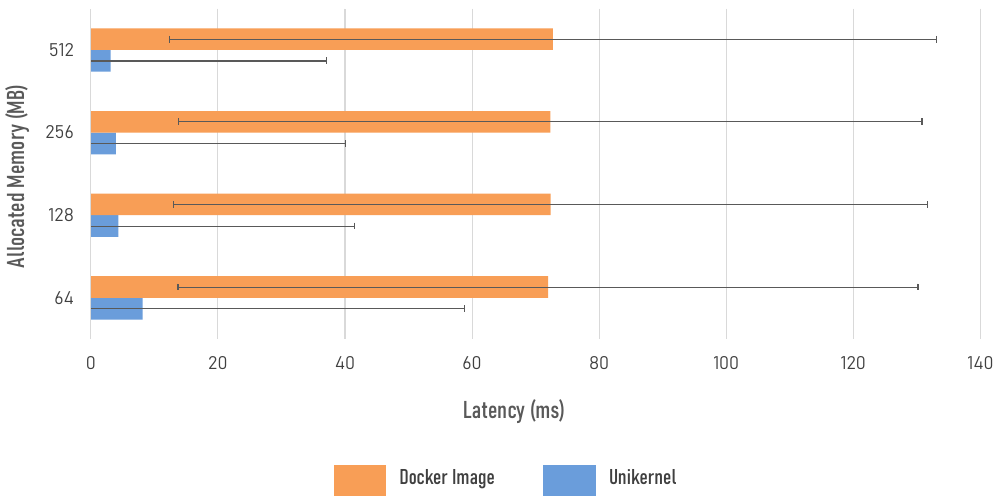}
\caption{Average Latency and Standard Deviation for the Go Application under Varying Memory Constraints (1 CPU). Error bars represent the standard deviation. Nanos exhibits both lower average latency and lower jitter.}
\label{fig:io_constrained_go_latency}
\end{figure}

The performance of the Go application in the constrained scenario revealed a strong reversal of the results seen in the previous tests, with the Nanos unikernel demonstrating a consistent advantage (Fig.~\ref{fig:io_constrained_go_throughput} and Fig.~\ref{fig:io_constrained_go_latency}).

While Docker's throughput remains stable with different memory allocation, the Nanos unikernel demonstrates better throughput overall. It delivered high and stable throughput at 512 MB and 256 MB (70,179.61 req/s and 70,410.22 req/s respectively), with a correspondingly low average latency of under 4ms. At these memory level, its performance is not limited by allocated memory. As the available memory was reduced to 128 MB, the throughput drops to  41,167.76 req/s while average latency reaches 4.31 ms. This indicates that at 128 MB, memory availability has become the primary bottleneck due to increased garbage-collection overhead.

The Docker container's performance remains stable around  6,000 req/s across all tested memory levels. The average latency was also consistently high at around 72 ms. This stability, despite varying memory allocations, suggests that CPU availability is the primary limiting factor in these performance results. This observation aligns with the deployment configuration of the minimal application used in the benchmark, which relied on default runtime settings. By default, Node.js utilizes a single-threaded \textit{event loop}, limiting its ability to leverage multiple CPU cores for concurrent request handling. In contrast, Go employs \textit{goroutines}, allowing it to utilize available CPU cores more effectively under concurrent workloads, resulting in improved scalability with increased CPU resources.

\begin{figure}[htbp]
\centering
\includegraphics[width=\linewidth]{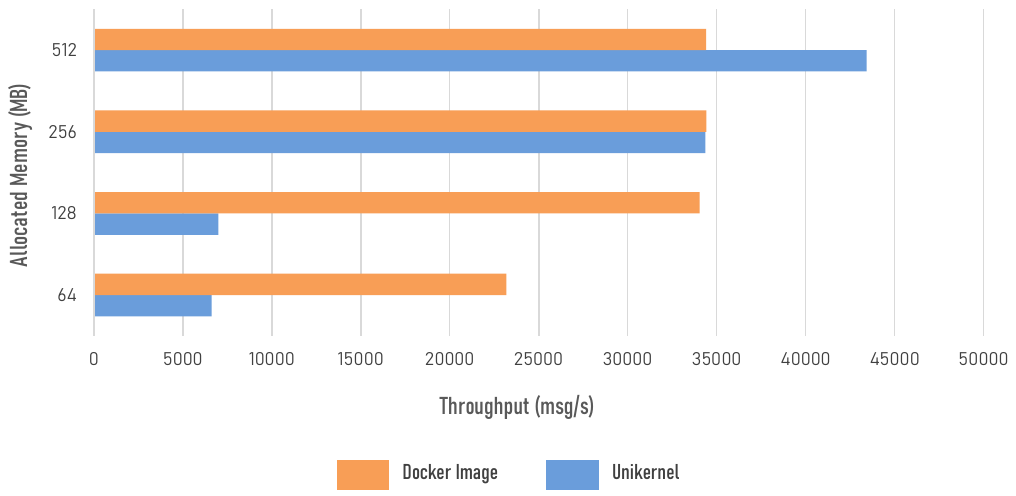}
\caption{I/O-Bound Throughput for the Node.js Application under Varying Memory Constraints (1 CPU). A performance crossover occurs at the 128 MB level, where Docker starts performing better.}
\label{fig:io_constrained_node_throughput}
\end{figure}

\begin{figure}[htbp]
\centering
\includegraphics[width=\linewidth]{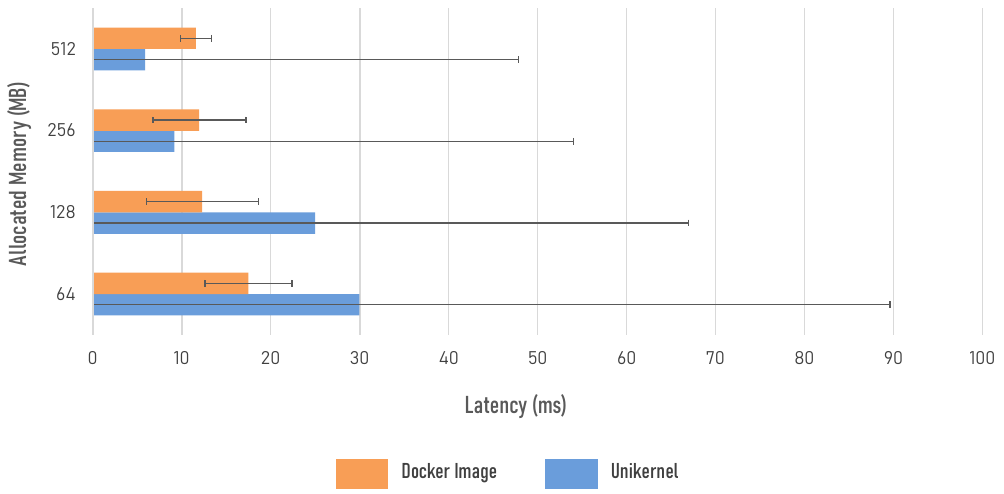}
\caption{Average Latency and Standard Deviation for the Node.js Application (1 CPU). Note the dramatic increase in latency and jitter for the unikernel at 128MB and below.}
\label{fig:io_constrained_node_latency}
\end{figure}

Unlike the straightforward results from the Go application, the Node.js benchmark reveals a different pattern,  where the container performs better than unikernel as memory becomes more constrained (Fig.~\ref{fig:io_constrained_node_throughput} and Fig.~\ref{fig:io_constrained_node_latency}). 

At higher memory allocations, the Nanos unikernel performs slightly better than Docker (43,425.20 req/s vs 34,411.89 req/s at 512 MB). The performance advantage for Nanos can be attributed to the absence of a full guest OS means fewer background processes and less scheduling overhead, allowing the Node.js event loop to operate with maximum efficiency. At 256 MB, the two platforms perform almost identically, suggesting this is the point where the memory advantage of Nanos is balanced by Docker. 

A performance reversal occurs at the 128 MB mark. The Nanos unikernel's throughput decreases sharply from 34,358.57 req/s (at 256 MB) to 6,992.61 req/s (at 128 MB), while the average latency increases significantly (from 9.17 ms to 25.00 ms). In stark contrast, the Docker container's performance remains stable and robust, maintaining ~34,052.28 req/s at 128 MB and only degrading moderately to ~23,175.36 req/s at 64 MB. 

This performance degradation is also observed with the Go application, although to a lesser extent (41.53\% drop vs. 79.65\% for Node.js). This indicates that the runtime environment itself is a contributing factor to the observed performance reduction. The Nanos unikernel, while lightweight and efficient under ample resources, uses custom memory allocators \cite{nanovms2025memory} and lacks the advanced memory management mechanisms of the Linux kernel. As a result, under severe memory constraints, workloads that rely on JIT compilation and dynamic heap management, such as Node.js, are particularly vulnerable, revealing the limitations of unikernels in handling memory-constrained environments for such workloads.

While Nanos unikernels can offer a performance advantage with more allocated memory, Docker containers provide a more stable and resilient platform for Node.js when memory is highly constrained. This finding is critical for architects, as it demonstrates that for certain application stacks, the memory overhead of a container is a worthwhile trade-off for the performance stability gained from a mature and well-understood operating system environment.

\subsection{Performance Evaluation: CPU-Bound Workload}

While many edge applications are I/O-bound, a significant and growing number of Industry 4.0 use cases involve intensive local data processing. These CPU-bound workloads include tasks such as data filtering, aggregation, real-time analytics, and executing machine learning inference models. This benchmark is designed to minimize the influence of the networking stack and isolate the raw computational efficiency of the application runtimes (Go vs. Node.js) on each platform (Nanos vs. Docker).

\subsubsection{Resource-Rich Scenario}

\begin{figure}[htbp]
\centering
\includegraphics[width=\linewidth]{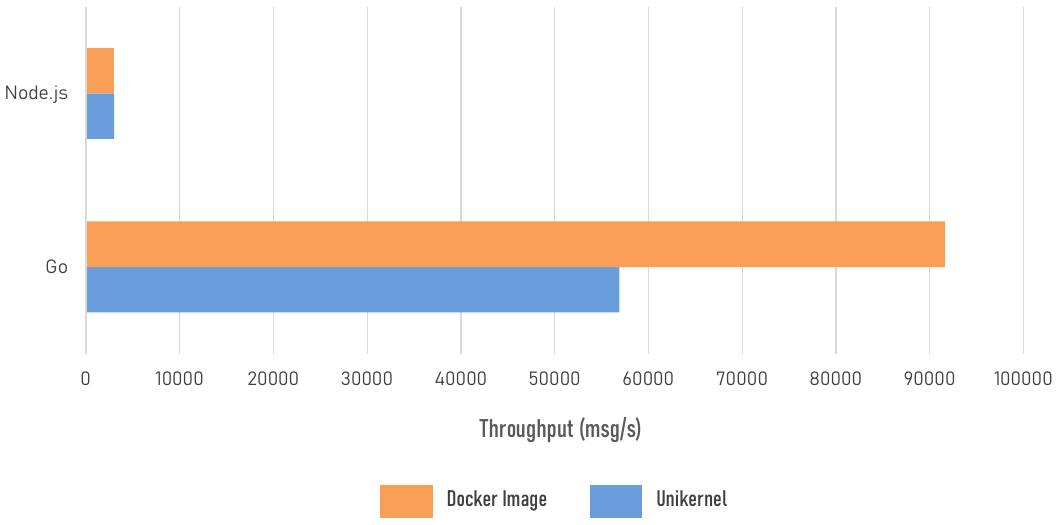}
\caption{CPU-Bound Throughput in an Unconstrained Scenario (8 vCPU, 2048 MB).}
\label{fig:cpu_throughput_unconstrained}
\end{figure}

This scenario measures the maximum parallel computation throughput when a generous resources are provided. Both platforms were allocated 8 CPU cores and 2048 MB of memory. The throughput results for this test are summarized in Fig.~\ref{fig:cpu_throughput_unconstrained}

A particularly notable finding is the stark performance difference between the Go and Node.js applications. As shown in Fig.~\ref{fig:cpu_throughput_unconstrained}, the Go application's throughput (91,655 req/s on Docker and 56,984 req/s on Nanos) up to 30 times higher than the Node.js application's (2,988.56 req/s and 3,027.46 req/s). This is a direct consequence of their different compilation models. The Go application, being pre-compiled to optimized native machine code (AOT), executes the computational loop with maximum efficiency. The Node.js application, which relies on the V8 engine's JIT compilation, incurs significant runtime overhead for this type of repetitive, CPU-intensive task. In addition, from the resource rich scenario of I/O-Bound workload, we also have observed the highly efficient multi-core utilization of Go. 

\subsubsection{Resource-Constrained Scenario}

This scenario evaluates the single-core computational efficiency of each platform under the same memory pressure tests as the I/O-bound workload (512 MB, 256 MB, 128 MB, 64 MB).

\begin{figure}[htbp]
\centering
\includegraphics[width=\linewidth]{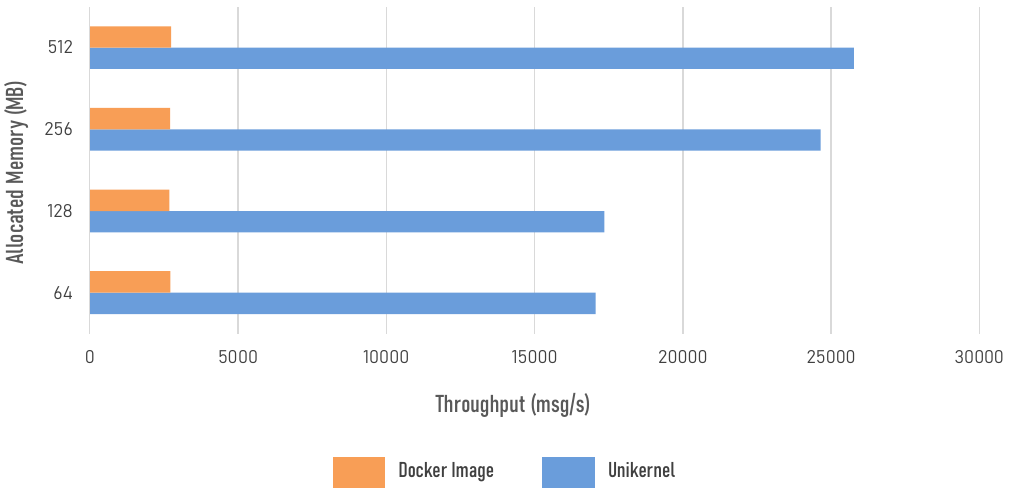}
\caption{CPU-Bound Throughput for the Go Application under Varying Memory Constraints (1 CPU). The Nanos unikernel's throughput is an order of magnitude higher than Docker's across all memory levels.}
\label{fig:cpu_constrained_go_throughput}
\end{figure}

\begin{figure}[htbp]
\centering
\includegraphics[width=\linewidth]{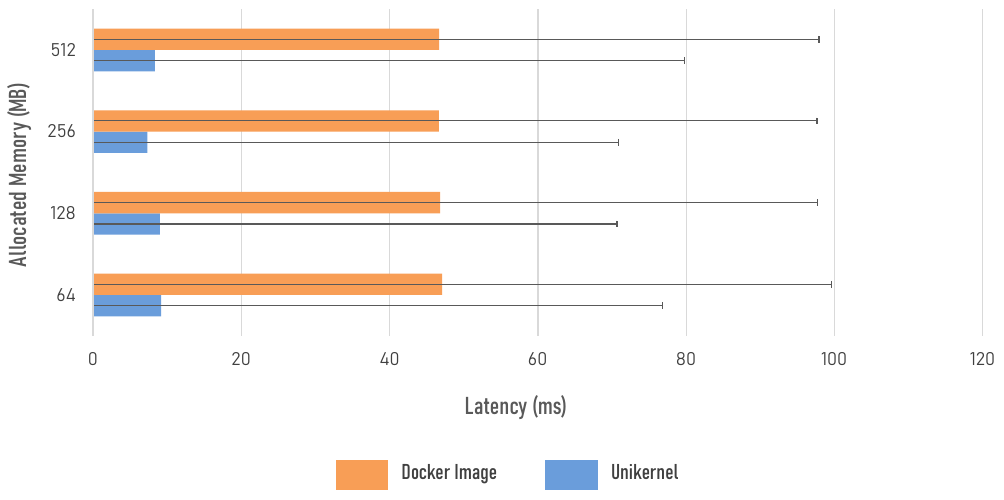}
\caption{Average Latency and Standard Deviation for the Go Application (1 CPU). The unikernel exhibits significantly lower average latency and less performance jitter compared to the container.}
\label{fig:cpu_constrained_go_latency}
\end{figure}

Results from the throughput test show a similar pattern as in the I/O-bound scenario, where the Go application performs better in Nanos compared to Docker at all memory levels as shown in Fig.~\ref{fig:cpu_constrained_go_throughput} and Fig.~\ref{fig:cpu_constrained_go_latency}. The unikernel, by design, has minimal system overhead. It does not run a separate guest operating system. When the Go application is executing its hashing loop, the single available CPU core is almost entirely dedicated to that task. There is minimal \textit{system noise} or background processes competing for CPU cycles. This high degree of efficiency allows the application to achieve its maximum possible computational output on the given hardware. The performance degradation observed when reducing memory from 256 MB to 128 MB is a predictable consequence of increased memory pressure, forcing the garbage collector to consume a larger portion of the limited CPU time.

The Docker container's performance is consistently low and flat at around 2,700 req/s. In this single-core environment, the overhead of running the containerized environment along with associated context switching and system management tasks, consumes a significant fraction of the available CPU cycles. This leaves less processing time available for the actual Go application workload. The system is bottlenecked not by the application itself, but by the inherent overhead of the platform architecture.

 \begin{figure}[htbp]
\centering
\includegraphics[width=\linewidth]{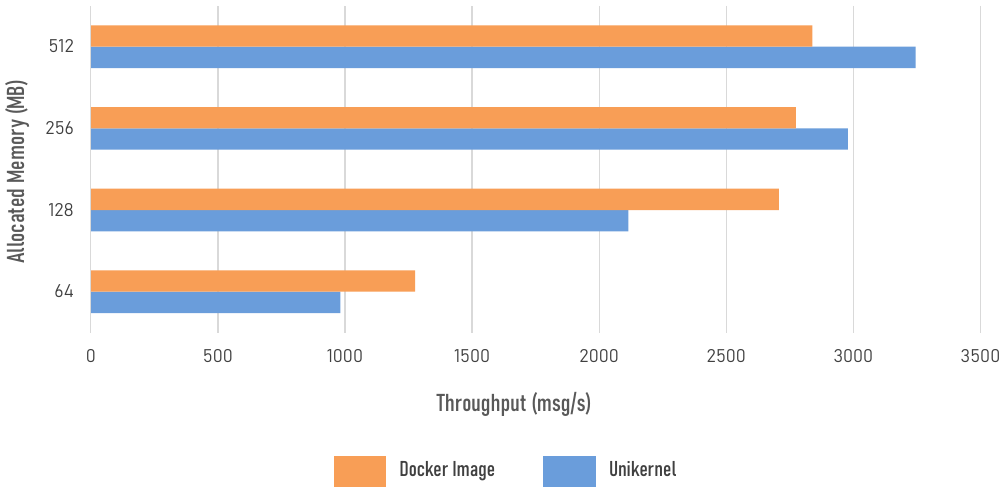}
\caption{CPU-Bound Throughput for the Node.js Application under Varying Memory Constraints (1 CPU). A performance crossover occurs below the 256 MB memory level, where Docker's stability provides a performance advantage.}
\label{fig:cpu_constrained_node_throughput}
\end{figure}

\begin{figure}[htbp]
\centering
\includegraphics[width=\linewidth]{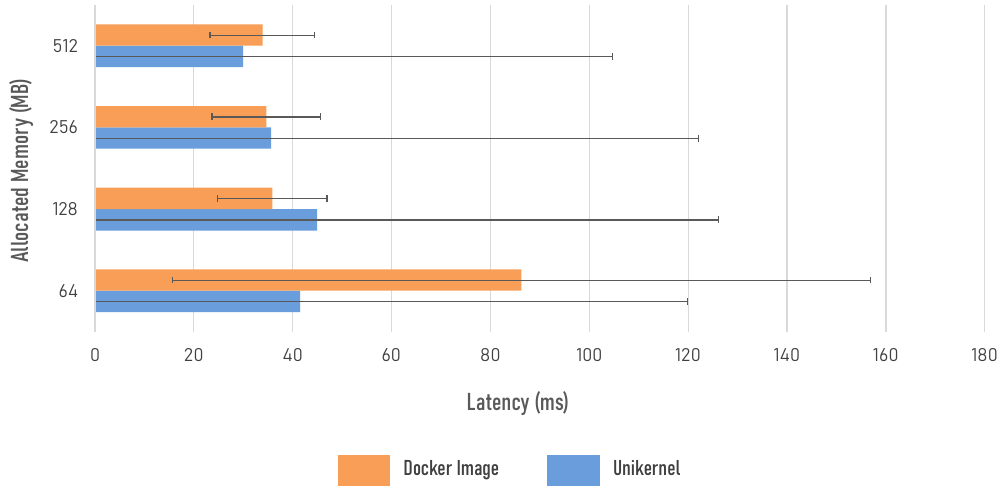}
\caption{Average Latency and Standard Deviation for the Node.js Application (1 CPU). The chart highlights the dramatic increase in latency and performance jitter for the unikernel at 128 MB and below.}
\label{fig:cpu_constrained_node_latency}
\end{figure}

With Node.js, the CPU-bound workloads demonstrate similar crossover pattern as previously discussed. Under sufficient memory, Nanos demonstrates a slight performance advantage, delivering 6.85-12.54\% higher throughput than Docker. In this case, runtime overhead is the main differentiator. As the allocated memory is reduced to 128 MB, the Docker container becomes the superior platform, outperforming the unikernel. The throughput of the Nanos unikernel reduces significantly (from nearly 2,977.81 req/s to 2114.19 req/s then to 980.90 req/s) as shown in Fig.~\ref{fig:cpu_constrained_node_throughput}. In contrast, the Docker container's performance degrades much more gracefully, maintaining a relatively stable throughput with significant degradation at 64 MB. Latency results (Fig.\ref{fig:cpu_constrained_node_latency}) confirm this trend, showing a significant increase at 128 MB and below.

\section{Discussion}

The evaluation results clarify how unikernels (Nanos) and containers (Docker) differ architecturally and impact application performance across workloads, languages, and resource constraints. As with microservice frameworks, there is no universal ``best'' platform; the optimal choice depends on the deployment context. From our benchmarks, we draw four key conclusions.

\subsection{Static Properties: Unikernel Agility}

Our results strongly support hypothesis \textbf{H1}, with unikernels achieving a \textit{12 ms startup} compared to \textit{158-219.5 ms for containers}, confirming the rapid instantiation advantages noted by Goethals et al. \cite{goethals2022functional}. For the industrial compute continuum, this is a critical enabler for resilience and elasticity at the edge. Services deployed as unikernels can be scaled up or recovered from failure in milliseconds, enabling a level of responsiveness that is difficult to achieve with containers for time-critical services.

Additionally, the \textit{2.4x reduction in image size} for Node.js applications aligns with Chen et al.\cite{chen2021evolving}, translating to lower storage and bandwidth needs, supporting fast Continuous Integration and Continuous Delivery/Deployment (CI/CD) cycles.

\subsection{I/O Throughput: Kernel vs.~Runtime Efficiency}

Our second hypothesis (\textbf{H2}) predicted that performance would be highly workload-dependent, and the I/O-bound results confirm that. In the resource-rich, multi-core scenario, Docker outperforms Nanos for Go applications. This performance gap is likely attributable not only to the deployment paradigm itself but also to the underlying networking model; Docker utilizes the host's highly optimized, kernel-level networking stack, while our Nanos setup defaults to user-space networking, which can introduce overhead in highly parallel I/O scenarios. This is a common architectural compromise in the unikernel space, where the goal of creating a minimal, lightweight artifact can lead to the use of less-performant, user-space libraries in place of a full kernel networking stack.

In contrast, Nanos outperforms Docker with the single-threaded Node.js application highlights a different bottleneck. Here, the limiting factor is not the network stack but the Node.js's event loop itself. In this context, the unikernel's minimal system overhead provides an advantage, allowing the event loop to run with less interference and greater efficiency.  It is also  worth mentioning that Docker, however, demonstrates lower jitter and fewer request drops, indicating its superior predictability under heavy loads. This demonstrates that for I/O workloads, the optimal paradigm choice depends on whether the bottleneck is the host's kernel-level networking layer  (favoring Docker/Linux) or the application runtime (favoring the lighter unikernel).

\subsection{CPU-Bound Performance: Specialization Advantage}

For CPU-bound workloads, the evaluation shows that Go outperforms Node.js by up to 30 times, demonstrating the benefits of AOT compilation over JIT-based execution in high-load scenarios. In multi-core environments, Docker offers higher throughput for Go due to the mature Linux scheduler efficiently distributing workloads across cores, while Node.js remains bottlenecked by its single-threaded model. In single-core, constrained environments, Nanos consistently outperforms Docker for Go, emphasizing the benefits of unikernel minimalism for compute-heavy tasks where every CPU cycle is critical. However, for Node.js, Docker again provides more stable performance under severe memory constraints, maintaining throughput where the unikernel struggles. The stability and optimizations of a standard Linux environment are critical to its performance under pressure, making Node.js + Docker a more resilient combination in severely constrained scenarios.

This finding validates the core value proposition of this type of specialized unikernel for specific, high-value industrial use cases. For analytic services at the extreme edge that perform intensive computation, such as real-time signal processing, data filtering, or cryptographic functions, a combination of unikernel and AOT-compiled languages offers a performance level that containers cannot match on constrained hardware.

\subsection{Crossover Point: Stability in Constrained Environments}

In resource-constrained edge environments, the programming language runtime determines the performance bottleneck. For Go, Nanos outperforms Docker by an order of magnitude across all memory levels in single-core tests. Although Nanos shows some degradation when reducing memory from 256 MB to 128 MB due to increased garbage collection, it still maintains a significant lead, demonstrating that the Nanos unikernel can dedicate nearly all CPU cycles to execution, unlike Docker, which incurs system overhead.

For Node.js, the pattern differs: Nanos holds an advantage at higher memory levels but suffers a sharp decline in throughput and increased latency below 128 MB, while Docker maintains stability. This reflects the Nanos unikernel's specific design, which lacks the advanced memory management mechanisms of the Linux kernel. This makes JIT-based workloads like Node.js vulnerable under severe constraints, partially confirming our \textbf{H2} hypothesis.

These findings imply that for JIT-based services (Node.js, Java, Python), Docker may be the more resilient and practical choice on constrained edge devices. Here, the container’s resource overhead is justified by the stability and predictability of the Linux kernel.  It is important, however, to scope this definition of stability: in this context, it refers to performance consistency under short-term memory pressure. Long-term operational reliability and fault tolerance were not assessed.

\section{Conclusion}

The empirical results and discussion demonstrate that selecting a deployment paradigm is not a binary decision but a complex architectural trade-off. The findings from this work enable us to move beyond theoretical arguments and propose a practical decision framework for engineers and architects implementing services for a compute continuum in Industry~4.0.

For time-sensitive analytics services performing intensive computation (e.g., analytics, signal processing) on highly resource-constrained edge devices, a unikernel running an AOT-compiled language such as Go or Rust is the superior choice. This combination yields an order-of-magnitude improvement in CPU efficiency, memory footprint, and startup latency. The minimal system overhead of unikernels is critical for extracting maximum performance from minimal hardware, making them ideal for extreme-edge deployments.

For services built with JIT-compiled runtimes (e.g., Node.js, Java, Python) on edge devices where memory is a primary constraint, Docker containers are the more resilient and stable choice. The mature memory management and process scheduling of the Linux kernel provide crucial stability, preventing the performance collapse observed with unikernels under severe memory pressure. Here, the architectural stability of Docker outweighs the raw resource savings offered by unikernels.

For high-throughput workloads on well-provisioned, multi-core edge servers or within cloud-based long-term analytics services, the optimal choice depends on the application's programming model. For parallel runtimes like Go or Rust, Docker containers are superior, as our results show that Docker utilizes the host’s highly optimized Linux kernel networking and scheduler. In contrast, for single-threaded runtimes where the runtime is the bottleneck, unikernels offer a slight performance advantage due to their minimal system overhead, allowing the application to operate with less interference and achieve marginally higher throughput under these conditions.

While our findings provide novel insights, they are bounded by the specific context of our experimental design, and we acknowledge three primary limitations that suggest directions for future work. First, our results are based on the Nanos unikernel, and it remains to be seen whether the observed performance crossover is a generalizable phenomenon or a consequence of Nanos’s particular memory management strategy; comparative evaluations involving other unikernel ecosystems such as Unikraft or MirageOS would be highly beneficial. Second, our I/O benchmarks reflect a comparison between a kernel-space networking model (Docker) and a user-space model (Nanos/ops), which introduces an additional variable; future studies could control for this by employing unikernels with kernel-bypass capabilities to better isolate the effects of the virtualization paradigm itself. Third, while synthetic workloads allowed us to maintain experimental control, validating these results against real-world industrial applications with mixed computational and I/O patterns could provide even more valuable insights.

In summary, this research demonstrates that the choice between containers and unikernels is a complex engineering trade-off, not a binary decision. The optimal paradigm is dictated by a detailed analysis of the application's workload, its runtime execution model (AOT vs. JIT), and the specific resource constraints of the target environment. This underscores the necessity of a systematic, context-aware approach to making architectural choices for edge computing in general and the industrial edge in particular.

\bibliographystyle{IEEEtran}   
\bibliography{references}      

\end{document}